\documentstyle[12pt]{article} 
\newlength{\dinwidth} 
\newlength{\dinmargin} 
\newcommand{\ba}{\begin{array}} 
\newcommand{\ea}{\end{array}} 
\newcommand{\be}{\begin{equation}} 
\newcommand{\ee}{\end{equation}} 
\newcommand{\bea}{\begin{eqnarray}} 
\newcommand{\eea}{\end{eqnarray}} 
 
\newcommand{\bx}{{\bf x}} 
 
\def\bee{\begin{eqnarray}} 
\def\eee{\end{eqnarray}} 
\def\be{\begin{equation}} 
\def\ee{\end{equation}}

\begin{document} 
\thispagestyle{empty} 
\addtocounter{page}{-1} 
\begin{flushright} 
UOSTP 99-105\\ 
SNUST 99-02 \\
{\tt hep-th/9902173}\\ 
{revised version} \\
\end{flushright} 
\vspace*{1.3cm} 
\centerline{\Large \bf Cosmic Holography\footnote{ 
Work supported in part by BK-21 Initiative Program and KRF International 
Collaboration Grant 1998-010-192. }} 
\vspace*{1.2cm} 
\centerline{\large \bf Dongsu Bak${}^1$ and Soo-Jong Rey${}^2$
} 
\vspace*{0.8cm} 
\centerline{\it Physics Department, 
University of Seoul, Seoul 130-743 Korea${}^1$} 
\vskip0.4cm 
\centerline{\it Physics Department \& Center for High-Energy Physics}
\vskip0.3cm
\centerline{\it Seoul National University, Seoul 151-742 Korea${}^2$} 
\vskip0.4cm 
\centerline{\tt dsbak@mach.uos.ac.kr, 
\quad sjrey@gravity.snu.ac.kr
} 
\vspace*{1.5cm} 
\centerline{\Large\bf abstract} 
\vspace*{0.5cm} 
A version of holographic principle for the cosmology is proposed, which
dictates that the particle entropy within the cosmological apparent 
horizon should not exceed the gravitational entropy associated with the 
apparent horizon. It is shown that, in the Friedmann-Robertson-Walker (FRW) 
cosmology, the open Universe  as well as a restricted class of flat 
cases are compatible with the principle, whereas closed Universe is not. 
It is also found that inflationary universe after the big-bang is incompatible
with the cosmic holography.
\vspace*{1.1cm} 
 
\baselineskip=17pt 
\newpage 
 
\setcounter{equation}{0} 
 
The holographic principle in quantum gravity is first suggested by 
't Hooft\cite{thooft} and, 
later, extended to string theory by Susskind\cite{susskind}.
The most radical part of the principle is that 
the degrees of freedom of a spatial region reside not in the bulk
but in the boundary. Further, the number of boundary degrees of freedom 
per Planck area should not exceed unity.
Recently, the holographic principle is applied to the standard 
cosmological context by Fischler and Susskind\cite{fischlersusskind}. 
This Fischler-Susskind version of cosmological holographic 
principle demands that the particle entrophy contained in a volume
of particle horizon should not exceed the area of the horizon in 
Planck units. The string cosmology is also tested by the 
Fischler-Susskind holographic principle\cite{srey}. In both cases, 
the matter contents as well as the spacetime geometry
of the universe are restricted by the holographic principle alone
and the results appear to be consistent with the recent measurement
of the redshift-to-distance relation and the theory of the large
scale structure formation\cite{perlmutter}.

In applying the holography in the cosmological context, several 
outstanding questions still remain unanswered. One of them is concerning
a natural choice of the holographic boundary. Fischler and Susskind have
chosen it to be the particle horizon, but it is not clear if the choice is
consistent with other physical principles. In this note, we will propose 
a simple choice of the boundary surface based 
on the concept of {\sl cosmological apparent horizon} that is a boundary 
hypersurface of an anti-trapped region and has a topology of ${\bf S}^2$. 
It turns out that there is natural gravitational entropy 
associated with the apparent horizon and the associated holographic principle 
demands that the particle entrophy inside the apparent horizon should not
exceed the apparent-horizon gravitational entropy. Moreover, the holography 
based on the apparent horizon obeys the {\sl first law} of thermodynamics, 
in sharp contrast to that based on the particle horizon.

We shall apply the proposed principle to the FRW cosmology and show that, 
in both the standard cosmology and the string cosmology, the open 
universe as well as restricted class of flat universe are compatible,
while the closed universe is not. We shall further show that the 
inflationary senario for the standard cosmology is not compatible with the 
cosmic holography.

\vskip0.5cm 
\sl Cosmological Apparent Horizon and Gravitational Entropy: \rm 
We shall consider the spatially homogeneous and isotropic 
universe described by the FRW metric,
\be 
ds^2= -dt^2+a^2(t){dr^2 \over 1 - k r^2} 
+ a^2(t)r^2 d \Omega^2_{d-1} \, , 
\label{frwmetric} 
\ee 
where $k = 0, -1, +1$ correspond to a flat, open or 
closed Universe  
respectively.
Using the spherical symmetry, the metric can be 
rewritten as
\be 
ds^2=h_{ab}dx^a dx^b   
+ {\tilde{r}}^2 (\bx) d \Omega^2_{d-1} \, , 
\label{twometric} 
\ee 
where $x^0=t$, $x^1=r$ and the two metric 
$h_{ab}={\rm diag}\left[-1,\, {a^2\over 1-k r^2}\right]$ is introduced. 
A dynamical  apparent horizon is defined by 
$h^{ab}\partial_a \tilde{r} \partial_b \tilde{r}=0$, which 
implies that the vector $\nabla\tilde{r}$ is 
 null (or degenerate) on the apparent horizon surface. The explicit 
evaluation of the condition reads
\be 
\tilde{r}_{\rm AH}={1\over \sqrt{H^2+{k\over a^2}}} \, , 
\label{nullcondition} 
\ee  
where $H=\dot{a}/a$ is the Hubble parameter.
The expansion $\theta_{\rm IN}$ 
($\theta_{{\rm OUT}}$) of the ingoing (outgoing) null 
geodesic congruence are
given by
\bee 
&&\theta_{\rm IN}=H -{1\over \tilde{r}} \sqrt{1-{k\tilde{r}^2 
\over a^2}} \nonumber\\
&&\theta_{\rm OUT}=H +{1\over \tilde{r}} \sqrt{1-{k\tilde{r}^2 
\over a^2}}\, . 
\label{expansion} 
\eee 
The region of  spherically symmetric spacetime is 
referred to as trapped (antitrapped) if the expansions of 
both in- and 
out-going null geodesics, normal to the spatial $d-1$ sphere 
with a radius $\tilde{r}$ 
centered 
at the origin, are negative (positive). The region will be called 
normal if ingoing rays have negative expansion but the 
outgoing rays have positive expansion. 
The region of $\tilde{r}> \left(H^2+{k\over a^2}\right)^{-{1\over 2}}$ is,  
then, antitrapped, whereas  the region of 
$\tilde{r}< \left(H^2+{k\over a^2}\right)^{-{1\over 2}}$
is normal (assuming $H > 0$). 
The boundary hypersurface of the antitrapped 
spacetime region is nothing but the apparent horizon 
surface. 
The ingoing rays outside the horizon actually propagate in the 
direction of the growing $\tilde{r}$, whereas the ingoing rays 
inside the horizon are moving toward the origin. 

In general, the radius of the apparant 
horizon, $\tilde{r}_{\rm AH}$, changes in time. 
But, for example, in the de Sitter universe where 
$a(t)=a_0 e^{Ht}$ with a constant $H$ and $k=0$, the 
apparent horizon, $\tilde{r}_{\rm AH}={1\over H}$, is constant 
in time and agrees with the cosmological event horizon of the 
de Sitter space. The antitrapped region outside of the 
apparent horizon in the de Sitter space can never be  seen
in the comoving observer located at the origin. However, 
in generic situation, the apparent horizon evolves in time 
and visiblity of the outside antitrapped region depends on
the time developement of the apparent horizon.
In case the $\tilde{r}_{\rm AH}$ becomes smaller in time,
the spatial region outside the horizon can never be seen.
On the other hand, if it grows, the spatial region outside
of the horizon at a given time may be observed at later time.
The situation here is reminiscent of what is happening 
with the black-hole 
apparent horizon. Namely, the trapped region can never be seen 
by an outside observer
if the horizon of the black grows by an infalling matter, 
while once-trapped region may become normal if the apparent  
horizon shrinks by an evaporation of the black hole in 
presence of the 
Hawking  radiation.

For the total energy inside a sphere of radius $\tilde{r}$,
we introduce an energy defined by
\be 
E\equiv {d(d-1) {\cal V}_d\over 16\pi G}\, \tilde{r}^{d-2}
(1-h^{ab}\partial_a \tilde{r} \partial_b \tilde{r})\, , 
\label{msenergy} 
\ee 
where ${\cal V}_d= {\pi^{d\over2}\over \Gamma \Bigl({d\over 2} +1\Bigr)}$ 
denotes the volume of the $d$-dimensional
unit ball. This is actually the direct $(d+1)$ dimensional 
generalization of the $(3\!+\!1)$ dimensional one given 
by Misner and Sharp\cite{misner}. It is interesting to note that
the energy surrounded by the apparent horizon is given by
$E\equiv {d(d-1) {\cal V}_d\over 16\pi G}\, \tilde{r}_{\rm AH}^{d-2}$,
which  agrees with the  
expression for the mass in the $d+1$ dimensional Schwarzschield
black hole once the apparent horizon is replaced by 
the event horizon
of the black hole. 

In terms of the energy-momentum tensor of matter $T^{ab}$ that is the 
projection of the $d+1$ energy-momentum tensor $T^{\alpha\beta}$ to the 
normal direction of the $d-1$ sphere, one may define the work density
by 
\be 
w\equiv  -{1\over 2} T^{ab}h_{ab}\, , 
\label{work} 
\ee 
and the energy-supply vector by
\be 
\psi_a \equiv  T_a\,^b \partial_b \tilde{r}  +w \partial_a\tilde{r}\, . 
\label{energysupply} 
\ee  
As noted in Ref.~\cite{hayward}, the work density at the apparent horizon 
may be viewed as  the work  done by the change of the apprent horizon and 
the energy-supply at the horizon is total energy flow through the 
apparent horizon.

The Einstein equation relates these quantities by
\be 
\nabla E = A \psi +w \nabla V\, . 
\label{firstlaw} 
\ee  
where $A= d{\cal V}_d \tilde{r}^{d-1}$ and $V={\cal V}_d \tilde{r}^d$. 
This equation may be  interpreted as {\sl unified first law}. The entropy
is associated with the energy-supply term, which in fact can be rewritten,
with again help of the Einstein equations,
as
\be 
A\psi = {\kappa\over 8\pi}\nabla A + \tilde{r}^{d-2}\nabla\Bigl(
{E\over \tilde{r}^{d-2}} 
\Bigr)\, , 
\label{entropy} 
\ee   
with the surface gravity $\kappa$ defined by
\be 
\kappa \equiv {1\over 2\sqrt{-h}}\partial_a \sqrt{-h} h^{ab}\partial_b
\tilde{r} \, . 
\label{surfacegravity} 
\ee   
At the apparent horizon, the last term in (\ref{entropy}) drops out and, 
then, the dynamic entropy of the gravity is identified with $S={A\over 4}$
that is a quarter of the area of the apparent horizon measured in the 
Planck unit. This is direct $(d+1)$ dimensional generalization of 
the definition of dynamic entropy introduced by Hayward for
the $(3+1)$ case\cite{hayward}. 
More precisely, the dynamic entropy associated 
with the apparent horizon is
\be 
S_{\rm AH}= {{\cal V}_d\over 4d}\tilde{r}^{d-1}_{\rm AP}\, . 
\label{apentropy} 
\ee   

We now apply the definition to the FRW universe dictated by
\bee 
 &&H^2 + {k \over a^2} = {16\pi\over d(d-1)}\,\rho \, ,   
\nonumber \\ 
&&{\ddot{a}\over a}= -{8\pi\over d(d-1)}[(d-2)\rho + dp] \, ,  
\label{equationsofmotion} 
\eee
with the energy-momentum conservation 
\be 
{d\over dt}(\rho a^d)+ p{d\over dt} a^d =0\, . 
\label{conservation} 
\ee  
The projected ($1+1$)-dimensional energy-momentum tensor 
for the FRW cosmology
reads 
\be 
T_{ab}={\rm diag}\left[\,\rho,\,\, {p\over 1-kr^2}\right]\, .
\label{frwenergytensor} 
\ee  
From (\ref{equationsofmotion}),
the Misner-Sharp energy can be evaluated, in terms of the 
matter density, as
\be 
E={d(d-1){\cal V}_d\over 16\pi}\left(H^2+{k\over a^2}\right)= 
{\cal V}_d \tilde{r}^d \rho\, ,
\label{frwenergy} 
\ee
which is the matter density multiplied by the flat-volume in 
$d$ spatial dimensions.  
One should note that the flat-volume is different 
from the spatial volume
 (i.e. $V_p=
d {\cal V}_d a^d \int^r_0 {dr\over\sqrt{1-kr^2}}$\,) of
 radius $\tilde{r}$ in case of the 
open or closed universe. This discrepancy appears 
due to the gravity contribution to the energy in addition 
to the matter contribution.

Having clarified the issue of the cosmological entropy, we now 
state a version of the cosmic holographic principle based on the
cosmological apparent horizon: {\sl 
the particle entropy inside the apparent horizon can never exceed
the apparent-horizon gravitational entropy.}

The main difference from the Fischler-Susskind version of the 
cosmological holography lies in the choice of the horizon; namely
in the Fischler-susskind proposal, the particle horizon and a quarter
of the associated area for the gravitational entropy is chosen for the 
holography.
In the cosmological context, there apeared two different kinds 
of horizon based on the light paths\cite{rindler}.  
The particle horizon that specifies the visible region for a comoving
observer  at time t,
is expressed as
\be 
\tilde{r}_{\rm PH}=a(t)G^{-1} \Bigl(\int^t_{t_I} {d t'\over a (t')}
\Bigr)\, ,
\label{particlehorizon} 
\ee
where $G(x)\equiv \int^x_0 {dy\over\sqrt{1-k y^2}}$ and $t_I$ 
represents the initial moment of the Universe. (In case the universe has
no beginning, $t_I=-\infty$.)
On the other hand,
the cosmological event horizon that specifies the boundary of 
the spatial region to be seen in the future  
by the comoving observer reads
\be 
\tilde{r}_{\rm EH}=a(t)G^{-1} \Bigl(\int_t^{t_F} {d t'\over a (t')}
\Bigr)\, ,
\label{eventhorizon} 
\ee
where $t_F$ is the final moment of the universe. This is contrasted to
 the fact that  the apparent 
horizon in (\ref{nullcondition}) does not refer to the 
initial or final moment where  our ability of physical 
description often breaks down. 
 
\vskip0.5cm 
{\sl Cosmic Holography Tested Against FRW Unverse}: 
The holographic principle may restrict the matter contents of
our universe because it involves the particle entropy of 
universe. Since matter contents are molding the geometry and 
evolution
of the Universe, the Universe 
itself conformed with the holography principle
may well belong to a  restricted class. 
The cosmic holography condition leads to an  
inequality    
\be 
\sigma {\rm Vol}_{\rm AH}(t) \le {A_{\rm AH}(t) \over 4 } .  
\label{thooft} 
\ee 
where ${\rm Vol}_{\rm AH}(t)=d{\rm V}_d 
\int^{r_{\rm AH}(t)}_0 dr{r^{d-1}\over 
\sqrt{1-kr^2}}$   
is the coordinate volume inside the apparent horizon.  The 
left side of (\ref{thooft}) is the total particle entropy 
inside the apparent horizon, where the coordinate entropy density
is constant in time.  In testing the condition to the FRW 
cosmology, we shall restrict ourselves to 
the cases of  matters with simple 
 equation of state $p=\gamma\rho$.

In the flat Universe ($k=0$), the condition is explicitly
\be 
{ 4\sigma \over d a^{d-1}(t)\dot{a}(t) } \le 1 .  
\label{flatcondition} 
\ee 
Since $a(t)= a_0 t^{{2\over d(1+\gamma)}}$ for $\gamma \neq -1$ and
$a(t)= a_0 e^{H_0 t}$ for $\gamma =-1$, one concludes that
the holography condition is satisfield all the time followed 
by the Planck time 
for $|\gamma|\le 1$ once it is satisfied at the Planck time, 
$t\sim t_P$. The matters with $|\gamma| > 1$  that 
are also inconsistent with special relativity,
 is not compatible with the cosmic holography condition.

Let us now turn to the 
case of open universe. For the discussion of open and 
closed universes, it is convenient to introduce the conformal time $\eta$,
\be 
\eta =\int^t {dt'\over a(t')}\, . 
\label{conformaltime} 
\ee 
The holography condition now reads 
\be 
{ 4\sigma \int^{\chi(\eta)}_0 d\xi \sinh^{d-1}\xi 
\over a^{d-1}(\eta) \sinh^{d-1}(\chi(\eta)) } \le 1 .  
\label{opencondition} 
\ee 
where we define $\chi(\eta)$ by $\sinh \chi(\eta)= r_{\rm AH}(\eta) $.
The solutions of the equation of motion (\ref{equationsofmotion}) are 
given by 
\be 
a(\eta)=a_0 \Bigl(\sinh |(K-1)\eta|\Bigr)^{1\over K-1}\, . 
\label{opensolution} 
\ee   
where $K= {d(1+\gamma)\over 2}$, $\eta\in (0,\,\infty)$ for $K-1>0$
and $\eta\in (-\infty,\, 0)$ for $K-1<0$.  Here the initial moment 
corresponds to $\eta=0$ for  $K-1>0$ and $-\infty$ for $K-1<0$. 
Using the definition of 
apparent horizon in (\ref{nullcondition}), one finds 
$\chi(\eta)= |(K\!-\!1)\eta|$. Using the explicit solution, 
one may easily show that, for  $\gamma \le 1$,
the holography condition is satisfied once satisfied at the 
initial moment around the Planck epoch.  For $\gamma >1$, the maximum 
of the left hand side
of (\ref{opencondition}) occurs at some finite time after 
initial moment and the
holography is respected if this this maximun satisfies the bound. 
But if one assumes 
that the holography bound is saturated at the Planck epoch,  
the case of $\gamma >1$
is rejected by the holography 
condition. So far the restrictions  on the matter 
contents 
due to our holography condition are not quite different from those of 
Fischler-Susskind holography. 
But we will see that there is some difference
in the case of the closed Universe.

The holography condition for the closed Universe ($k=1$) 
is 
\be 
{ 4\sigma \int^{\chi(\eta)}_0 d\xi \sin^{d-1}\xi 
\over a^{d-1}(\eta) \sin^{d-1}(\chi(\eta)) } \le 1 .  
\label{closedcondition} 
\ee 
where we define $\chi(\eta)$ by $\sin\chi(\eta)= r_{\rm AH}(\eta) $.
The solutions of the equation of motion (\ref{equationsofmotion}) are 
given by 
\be 
a(\eta)=a_0 \Bigl(\sin|(K\!-\!1)\eta|\Bigr)^{1\over K-1}\, . 
\label{closedsolution} 
\ee   
where 
$\eta\in \left(0,\, {1\over|K-1|}\right)$. 
For the Universe 
with $ K-1 <0$ 
(i.e. $\gamma < {2\over d} -1$) begins 
with infinite scale factor $a(\eta)$, and we shall 
not  discuss  these cases since it is clear 
that our Universe begins with small
scale factor from the observational ground.  Noting 
$r_{\rm AH}(\eta)=\sin(|(K\!-\!1)|\eta)$, one obtains 
that $\chi(\eta)= |K\!-\!1|\eta$.

Then one finds that the 
universe with $\gamma > {2\over d} -1$ are not compatible with 
the cosmic holography condition. Namely, even if one may satisfy the
bound at an  initial moment, it is badly violated before reaching the big 
crunch. This is because the term $ \int^{\chi(\eta)}_0 d\xi \sin^{d-1}\xi$
 monotonically grow with time. The situation is worsened if one assumes
that the bound saturates at the Planck time scale. 

In the Fischler-Susskind case, a careful analysis shows that
the disfavored region by the holography condition is only for 
$\gamma \le  {4\over d} -1$. With $d=3$, the universe with 
the matter of ${4\over d}-1={1\over 3}$ corresponds 
to radiation-dominated Universe, which is their marginal bound. 
On the contrary, the cosmic holography condition seems to disfavor any 
closed Universe, cleary an over-restrictive condition.

We now consider inflationary model of our $d=3$ Universe.
As illustrated in Ref.~\cite{fischlersusskind}, the 
particle entropy-area ratio $\alpha (t)\equiv 
{4\sigma {\rm Vol}_{\rm AH}(t)\over A_{\rm AH}(t)}$ at the time of 
decoupling is\footnote{The estimation used in the 
Ref.~\cite{fischlersusskind} relies upon the particle-horizon based
holography, but, in case of standard cosmology, the particle horizon 
differs from apparent horizon by a constant factor of order one.
Thus the estimation is still valid for our version of the holography.}   
\be 
\alpha (t_D)\sim 10^{-28},  
\label{decoupling} 
\ee 
where the decoupling time is $t_D\sim 10^{56}$ in Planck units.
Since $\alpha (t)$ is proportional to $t^{-{1\over 2}}$ 
during the radiation dominant era, the expresion of the ratio 
in the era reads
\be 
\alpha (t)= 10^{-28}\left[{t_D\over t}\right]^{1\over 2}.  
\label{radiation} 
\ee
This show that $\alpha(t) \le 1$ for all the time later Planck time
in case  Universe starts off as a radiation-dominant Universe after
the Planck time.
However, if one assume the inflationary periods 
after the Planck epoch and an exit to the radiation-dominant Universe,
the above conclusion drastically changed. To illustrate this, let us 
note that in the de Sitter phase, the $\alpha (t)$ scales like 
$e^{-(d+1)H t}= e^{-4Ht}$ with constant $H$. 
For example, the inflationary  factor $P\equiv e^{Ht_E}/e^{H t_B}
\sim 10^{100000}$ for the chaotic inflationary senario, 
where $t_E$ and $t_B$ denote respectively 
the exit time and the beginning of the inflation, 
is obtained from the theory of galaxy 
formation\cite{linde}. This implies that the $\alpha(t_B)$
is bigger than $\alpha(t_E)$ by a factor $10^{400000}$. The cosmic
holography condition is clearly incompatible with this result. 
Furthermore, any models where $\alpha(t)$
scales $t^{C}$ with $C < -{1\over 2}$ after Planck epoch, 
violates the holographic principle, so any typical 
post-big-bang inflationary models that solve the 
the flatness problem by an amplification of the scale factor 
appear to be incompatible with the holography. 
Does only the pre-big-bang super-inflationary 
senario\cite{veneziano} survive out of the restrictive holography condition 
or, otherwise, should one resort to the regularity of
the Planck epoch to solve the flatness and the horizon problem?

One is thus led to conclude that the holography based on apparent horizon,
despite its aesthetically simple and appealing features such as compatibility 
with the first-law of thermodynamics, are not totally satisfactory when 
applied to cosmology. Nevertheless, we trust that our proposal based on 
apparent horizon bears an important core of truth {\sl provided} further
specification is supplied on the holographic surfaces and bounded regions 
therein. 
After the present work has appeared, Bousso has made an interesting proposal 
\cite{bousso} concerning a covariant entropy bound valid for all surfaces 
in all physical spacetimes. In particular, Bousso has established that
Bekenstein bound holds if the surface permits a {\sl complete, 
future-directed, 
ingoing} null geodesic congruences. This singles out the apparent horizon
as the largest admissible holographic surface, thus confirming our proposal. 
Bousso's proposal has also remedied nicely difficulties we have posed in the
present paper. 
Most relevantly, it was shown that the holography bound is satisfied
only if the surface obeys the completeness condition and, when
dealing with closed universe, the entropy ought to be interpreted as that on 
null geodesic congruences directed toward the smaller part of the universe.   

Before closing, we would like to make a brief comments on the string 
cosmology. If one applies the new cosmic holographic principle to 
the string cosmology model considered in Ref.\cite{srey},
it is straghtforward to show that it selects out open Universe
together with  flat universes 
with $|\gamma|\le {1\over \sqrt{d}}$, while the closed universes are 
ruled out. Here one must note that the apparent horizon
should be defined with the Einstein frame metric instead of the 
string-frame metric because the gravitational entropy is associated 
with the Einstein frame metric although the relevant physics 
should be independent of the choice of description. In view of Bousso's
proposal \cite{bousso}, the above results should be interpreted as the 
condition on completeness of the apparent horizon.  
The above results are true both the pre-big-bang branch
and the post-big-bang branch, so that the time reversal 
symmetry is respected by the cosmic holography.
In the analysis\cite{srey} using the Fischler-Susskind version,
any flat Universes with matters are ruled out by the holography 
condition. This difference stems from the fact that the particle 
horizon in the pre-big-bang cosmology is infinite whereas the 
apparent horizon is finite in the pre-big-bang branch.


\end{document}